\documentstyle[pra,aps,preprint]{revtex}

\tighten

\def\ra{\rangle}
\def\la{\langle}
\def\vbar{\arrowvert}
\begin{document}
\title{Entropy decrease in Quantum Zeno Effect}
\author{Arun K. Pati$^{(1)}$}
\address{ School of Informatics, Dean street, University of Wales, Bangor LL 57 1UT, UK}

\address{(1) Theoretical Physics Division, 5th Floor, Central Complex,}
\address{Bhabha Atomic Research Centre, Mumbai - 400 085, INDIA.}
\date{\today}
\maketitle

\begin{abstract}
If a measurement process is regarded as an irreversible process, then by
Second law of thermodynamics the entropy should increase after any
measurement process. By the same spirit a quantum system undergoing
repeated measurement should show strong irreversibility leading to
entropy production. On the contrary we show that in quantum Zeno effect
setting the entropy of a quantum system decreases and goes to zero after
a large number of measurements. We discuss the entropy change under
continuous measurement model and show that entropy can decrease if we use
a more accurate measuring apparatus.
\end{abstract}

\vskip .4cm

email:akpati@sees.bangor.ac.uk

\vskip 1cm

Since its inception the concept of the entropy of a state has played an
important role in understanding enigmas of physics in diverse areas such
as thermodynamics, quantum mechanics, information theory and recently in
the area of quantum information processing. Traditionally one associates
some sort of ``disorder'' with entropy of a physical system. The hallmark
of second law of thermodynamics is that the entropy of an isolated system
increases with time or remains constant. However, it increases only when a
system undergoes an irreversible process (which in turn attributes an arrow
of time). In terms of negentropy this should decrease for an irreversible
dynamical changes in the system. Given a quantum system, if it is left
undisturbed and allowed to evolve unitarily then the entropy remains constant
with respect to a given preparation. This is rigidly connected with the
principle of linearity of time evolution of quantum system, because one
\cite{aper} could show that if the time evolution equation is non-linear
then entropy of a mixture of quantum system could spontaneously decrease in
a closed system.

But what about the entropy of a quantum system under observation? Since
observation(measurement) on a quantum system is an irreversible process one
would say that the entropy should increase as was first discussed by von
Neumann\cite{vn} in formulating his measurement theory. The entropy change
occuring in an isolated quantum system without measurement and entropy
increase due to measurement are quite different as expounded in a lucid book
by Brillouin\cite{br}. In the later case some amount of information can be
obtained as measurement involves an experiment with system and apparatus.
Thus, the information gain about a physical system by measuring a complete
set of commuting observables must be paid for in negative entropy(negentropy).
This would mean that if we do repeated measurements on a quantum system we
would gain more and more information and the quantum system should show a
strong irreversibility leading to increase of entropy. In other words the
above reasoning would inevitably lead us to say that frequent observation
on a quantum system system should increase the entropy of the system.

In this paper we investigate the entropy of a quantum system under repeated
observation. To be specific we look for the entropy change when the unitary
evolution of a quantum system is interrupted by sequence of measurements
(we call such a dynamics quantum Zeno
dynamics (QZD)). This would also answer the question:how does entropy change
when we tend to know more and more about the evolution of a quantum system.
Contary to aforesaid paragraph, we find that the repeated measurements on
a quantum system tends to decrease the entropy of a quantum system and under
continuous observation the entropy goes to zero. This can be proved within
von Neumann's collapse mechanism and unitary time evolution. Also, the same
can be proved using a continuous measurement model where the effective
evolution is non-unitary. In the sequel, we touch questions like does the
entropy depend upon the precision with which an apparatus measures the
observable of the system under study. With better measuring apparatus can
we be able to get much detailed information about the system? Does this
decrease the entropy? Since there is considerable interest in the theoretical
and experimental issues of quantum Zeno effect(QZE) we believe that the
present results will be of importance in answering certain subtle issues
like entropy and information under repeated observation.

The quantum Zeno effect (QZE) was originally discovered for an unstable
quantum system by Misra and Sudarshan \cite{ms}. For a coherent system QZE
says that if we prepare the system initially in an eigenstate of some
observable and repeatedly disturbes the unitary evolution of the system by
successive measurements, then the quantum transition to other states can
be completely suppressed.
Recently, Itano et al \cite{ita} have carried out an experiment to test the QZE
following a proposal of Cook \cite{ck}. This experiment gave rise to debates
over the fundamental issues of quantum theory which have been discussed by
several authors \cite{spmn,hnmn,hard} and also by the present author
\cite{akp}. In a continuing debate the
present author and Lawande \cite{apl} has questioned the necessity of
Schr\"odinger time evolution and argued that QZE could be observed in
non-linear quantum systems. Let us consider a quantum system
which has been prepared in the eigenstate of some observable $A$ that we are
interested in measuring. The observable $A$ has a discrete spectrum $\{a_n\}$
and a complete set of eigenstates $\{ \vbar \psi_n \ra\}$. In the absence
of any measurement the system at
a later time $t$ will make transition to other states under the
action of some unitary operator and the probabilities are distributed
according to $ p_n = \vbar c_n \vbar ^2$. Thus the state at time $t=0$
evolves to a state at time $t$ given by

\begin{equation}
\vbar \psi(t)\ra  = U(t) \vbar \psi(0) \ra = \sum_n c_n(t) \vbar \psi_n \ra 
\end{equation}

One can associate an entropy of the quantum system given these probability
distributions$\{p_1,p_2,...p_n\}$ as given by

\begin{equation}
 S(t) = -\sum_n p_n(t) \log p_n(t) 
\end{equation}
which is called Shannon entropy in information theory. This entropy depends
on the initial preparation stage of a quantum system. We will discuss the effect
of repeated measurements on the Shannon entropy with respect to a given preparation
stage. Since we have prepared our system in the eigenstate of some observable
$A$ the amplitudes can be written as

\begin{equation}
 c_n(t) = \la \psi_n \vbar e^{-iHt/ \hbar} \vbar \psi_n \ra  =
 \sum_m e^{-iE_m t/ \hbar}P_{nm}
\end{equation}
where $P_{nm} = \vbar \la \psi_n \vbar \phi_m \ra \vbar^2$ and $\{ \vbar \phi_m \ra \}$ being
the basis in which $H$ diagonalises. The matrix $P_{nm}$ are transition
 probability matrix elements some times called ``doubly stochastic matrix''
 which satsify $\sum_n P_{nm} = I = \sum_m P_{nm}$. Now the probability
 distributions are given by

\begin{equation}
 p_n(t) = \sum_{mk} e^{-i(E_m - E_k)t/ \hbar}P_{nm} P_{nk}.
\end{equation}

Therefore, the Shannon entropy corresponding to the preparation of the
system in the eigenstate of $A$ leads to the expression

\begin{equation}
S(t) = -\sum_n \biggl[ \sum_{mk} \cos \omega_{mk}t P_{nm} P_{nk} \log(\sum_{mk} \cos \omega_{mk}t P_{nm} P_{nk}) \biggr]
\end{equation}
where $\omega_{mk} = (E_m - E_k)t / \hbar$ is the transition frequency
between two energy levels and we have dropped the imaginary part of (3)
because $p_n(t)$ are real quantities. On the other hand if we could prepare
the system in the eigenstate of the Hamiltonian then the initial and final
entropy remain the same.  But this preparation stage is not interesting from
the standpoint of QZE becuase the probability of finding the system in the
nth eigenstate is always unity irrespective of repeated measurements. Therefore,
it is essential that we should prepare our system in eigenstate of some observable
which does not commute with the Hamiltonian of the system\cite{akp}.

 We investigate the Shannon entropy of the system
when  the unitary evolution during
the  time  interval  $[0,T]$ is interrupted by von Neumann measurements
such  that one performs a series of
measurements at times $\tau,  2\tau  \ldots  (N-1)\tau,  N\tau  =  T$.
During  the  short time interval  $[0,\tau]$ the system
evolves unitarily. The sequence of measurements that are  carried
out  are  idealised  to  be  discrete  and  instantaneous.

After  performing  a  von  Neumann  measurement  at  time  $\tau$
the probability of finding the system in the $n$th state is given by

\begin{equation}
 p_n(\tau) = 1 - {\tau^2 \over 2} \sum_{mk} {\omega_{mk}}^2 P_{nm} P_{nk}
\end{equation}

When the system undergoes repeated measurements $N$ number of times the
probability of finding the system in the $n$th state is given by

\begin{equation}
 p_n(T) = [p_n(\tau)^N] = \bigl(1 - {T^2 \over 2N} \sum_{mk} {\omega_{mk}}^2 P_{nm} P_{nk} \bigr)^N
\end{equation}

Thus the probabilities are distributed according to above rule after $N$
number of measurements and hence the Shannon entropy of the system is given by

\begin{equation}
 S_n(T) =  {T^2 \over 2N} \sum_n \biggl[\sum_{mk} {\omega_{mk}}^2 P_{nm} P_{nk} exp(- {T^2 \over 2N}\sum_{mk} {\omega_{mk}}^2 P_{nm} P_{nk}) \biggr]
\end{equation}
showing a  clear dependence on the number of measurements performed on the quantum
system. In the above expression we have used the large $N$ limit of the probability
distributions. From (5) one can conclude that as the number of measrements tend
to infinity the Shannon entropy of the system goes to zero. This is contrary to our
intution that large number of meaurements should be a signature of strong
irreversibility leading to entropy increase. Rather, we find that the entropy
of the system decreases (as it starts from a finite non-zero value and goes
to zero) when we tend to know more and more about the evolution of a quantum
 system.

The same result can also be proved within a continuous measurement
model \cite{ono,kula} which has been often invoked to simulate the quantum Zeno
effect without using von Neumann's projection postulate. These models are
different
from the continuous measurement models described by stochastic equation for
 the quantum sysytem and quantum trajectory approach. The measurement
 process in the above case is  described by a non-unitary evolution equation. This
 is done usually by taking an effective Hamiltonian which is non-Hermitian in
 nature. This kind of model has been very useful in proving new results and
 predicting new quantum effect such as quantum Zeno Phase effect (QZPE)
 \cite{pl,apsl}.
 For details one can refer to \cite{ono,pl}. We briefly
 recall that the evolution equation for a quantum system undergoing
 continuous  measurement of some observable $A$ is given by

\begin{equation}
i \hbar {d \vbar \psi(t) \ra \over dt} = \biggl[H - i \hbar f g \bigl({(A - a)
\over \Delta a} \bigr)
\biggr]  \vbar \psi(t)>
\end{equation}
where $H$ is the free Hamiltonian, $a$ is the result of the measurement that
the apparatus reads, $\Delta a$ is the the accuracy of the measurement
process and $f$ is the rate of information gain on the observable of the system.
The function $g({(A - a) \over \Delta a})$ takes care the interaction between the
system and apparatus in a parameter dependent way. The physical basis for such a phenomenological equation has been given on
restricted path integral approach \cite{onof} and also heuristically in \cite{kula}.

   Let us prepare our system as before in the eigenstate of the observable
   $A$ and follow the continuous evolution equation, then the time evolution
   can be given by

\begin{equation}
\vbar \psi(t) \ra = \exp[-i (H - i \hbar f g({(A - a) \over \Delta a}) )t / \hbar] \vbar \psi_n \ra.
\end{equation}
Now the probability amplitude in the nth state would be given by

\begin{equation}
c_n(t) = \la \psi_n \vbar e^{-i (H - i \hbar f u(A;a,\Delta a))t / \hbar} \vbar \psi_n \ra.
\end{equation}
where $u(A;a,\Delta a) = g({(A - a) \over \Delta a})$. The above equation
is in general difficult to simplify because the observable
$A$ does not commute with the Hamiltonian. But we can use
Campbell-Beker-Hausdorff formula to simplify it to some extent. Without loss
of generality for our purpose we assume that the commutator of $H$ and $A$
commutes with $H$ and $A$. In that case we can express the amplitudes as

\begin{equation}
c_n(t) = e^{- f g({(a_n - a) \over \Delta a})t} \la \psi_n \vbar e^{-i H t}e^{ - i/2 f t [H, u(A)]} \vbar \psi_n \ra.
\end{equation}
Therefore, the probabilities are now distributed according to

\begin{equation}
p_n(t) = e^{- 2f g({(a_n - a) \over \Delta a})t} \vbar V_{nn} \vbar ^2 .
\end{equation}
where $V_{nn}=  \la \psi_n \vbar e^{-i H t}e^{ - i/2 f t [H, u(A)]} \vbar
\psi_n \ra $ is the
matrix element involving all other operators. With this distribution we can see
that Shannon entropy is given by

\begin{equation}
S(t) = 2f \sum_n u_n e^{- 2f u_n } ÝV_{nn}Ý^2 - \sum_n e^{- 2f u_n } ÝV_{nn}Ý^2 \log ÝV_{nn}Ý^2.
\end{equation}
From the above expression one can clearly see that when we obtain more
information about the observable, i.e., in the limit of high frequency of
measurement the Shannon entropy goes to zero. It is interesting to note that if we
consider an observable which commutes with Hamiltonian then the term $ÝV_{nn}Ý^2$ is
unity and the entropy takes a simple form $S = 2f \sum_n u_n e^{-2f u_n} $,
which also goes to zero in the limit of high frequency of measurements.
Note that in the continuous measurement model we can talk of a commuting
observable with Hamiltonian and can still have an interesting physical
situation.

This model also provides answer to the question: Does the entropy depend on
the accuracy of the measuring apparatus? Does it decrease with increasing
the accuracy of the device? The answer is yes. As we can see when the accuracy
of the device increases the term $g(A;a,\Delta a)$ tends to infinite and then the
entropy again goes to zero. This is possible because the function $g(A;a,\Delta a)$
is a positive function of its argument, i.e., $g(x) \ge 0, g(0) = 0$. Generally,
it is assumed that $g(x) = x^2$ which gives a gussian type function in the
time evolution operator.

To conclude this paper we have shown that the Shannon entropy of a quantum
system decreases, and, in fact, goes to zero when it is interrupted by
a large sequence of measurements of the von Neumann type. The same result
is proved within a continuous measurement model. This is somewhat counter
intutive that measurement which is supposed to increase the entropy of the system,
the repeated measurements do the opposite.  This result raises several
questions which are related to quantum measurement theory, second law of
thermodynamics and the nature of entropy itself. Is it that  a quantum system
which has undergone several measurements in the past occupies a lowest
favourable state (since entropy is minimum). Is it that all the well
organised things that we see arround are results of continuous measurements
that our whole universe is undergoing? We hope that the new effect of entropy
decrease in the quantum zeno dynamics setting may be another way to achive
lowest entropy states.

I acknowledge the financial support from EPSRC. I thank R. Onofrio for
useful comments on this paper.

\renewcommand{\baselinestretch}{1}

\end{document}